\documentclass[%
 reprint,
 amsmath,amssymb,
 aps,
pra,
]{revtex4-1}

\usepackage{graphicx}
\usepackage{dcolumn}
\usepackage{bm}

\usepackage[capitalise]{cleveref}
\usepackage{xcolor}
\usepackage{physics}
\usepackage{soul}

\newcommand{\kin}{\kappa_\mathrm{in}}

\newcommand{\F}{\mathcal{F}}
\newcommand{\W}{\mathcal{W}}
\newcommand{\FSR}{\omega_\mathrm{FSR}}
\newcommand{\Trt}{T_\mathrm{rt}}
\renewcommand{\d}{\mathrm{d}}
\newcommand{\e}{\mathrm{e}}

\newcommand{\Tin}{\sigma_\mathrm{in}}
\newcommand{\Tout}{\sigma_\mathrm{out}}

\begin{document}

\preprint{APS/123-QED}

\title{Temporal stretching of single-photon pulses by intra-cavity frequency conversion}

\author{Mathias L. H. Korsgaard}
 \email{mathias.korsgaard@nbi.ku.dk}
\author{Eva M. González-Ruiz}%
\author{Anders S. Sørensen}
\affiliation{%
Center for Hybrid Quantum Networks (Hy-Q), Niels Bohr Institute, University of Copenhagen, Jagtvej 155A, 2200 Copenhagen N, Denmark
}%

\date{\today}

\begin{abstract}
Spectral bandwidth and frequency mismatch between quantum light emitters and quantum components is a common problem in the realization of hybrid quantum information networks.
Here we propose a simultaneous solution to both of these problems by using a free-space cavity with an embedded nonlinear crystal that, through frequency conversion, facilitates perfect coupling into a cavity mode.
The slow leakage of the cavity effectively stretches the photon pulse, resulting in a much longer photon pulse at a different wavelength.
We show that the amount of stretching is strictly limited by the finesse and length of the cavity.
Including realistic internal losses in the cavity, we show that it is possible to stretch a photon by at least two orders of magnitude with high efficiency.
The proposed setup can thus constitute a valuable addition to the toolbox of quantum information technologies.

\end{abstract}

\maketitle

\section{Introduction}

Single-photon pulses play a central role in hybrid quantum networks by acting as efficient low-loss carriers of quantum information between different physical quantum nodes~\cite{cirac_quantum_1997,divincenzo_physical_2000,kimble_quantum_2008}.
However, state of the art quantum technologies, such as quantum memories~\cite{bussieres_prospective_2013,afzelius_quantum_2015,heshami_quantum_2016}, transducers~\cite{engelhardt_optimal_2022,lauk_perspectives_2020}, mechanical oscillators~\cite{anetsberger_ultralow-dissipation_2008,rossi_measurement-based_2018}, and quantum dot single photon emitters~\cite{uppu_scalable_2020} often operate on timescales that differ by orders of magnitude and at different wavelengths.
This creates a fundamental temporal and spectral mismatch.
As an example, quantum dot single-photon sources have a bandwidth exceeding 150 MHz and typically operates at a wavelength around 950\,nm~\cite{kuhlmann_charge_2013,uppu_quantum-dot-based_2021}, while a rare earth doped solid-state quantum memory based on atomic frequency combs have a bandwidth in the range of a few MHz and a center wavelength of 606\,nm~\cite{hanni_heralded_2025,teller_solid-state_2025}.
Nonlinear frequency conversion in optical fibers combined with dispersion engineering has been used to simultaneously achieve frequency translation and pulse-reshaping~\cite{allgaier_highly_2017}, although the resulting bandwidths are still orders of magnitude away from, e.g. the bandwidth of typical quantum memories~\cite{bussieres_prospective_2013,afzelius_multimode_2009,heshami_quantum_2016,afzelius_quantum_2015,cho_highly_2016,vernaz-gris_highly-efficient_2018,hanni_heralded_2025}.
Alternatively, recent developments in photon storage with atomic ensembles have achieved arbitrary pulse shaping covering several orders of magnitude in bandwidth~\cite{mohan_deterministic_2026}, however, this does not include frequency conversion.

\begin{figure}[b]
    \centering
    \includegraphics[width=1\linewidth]{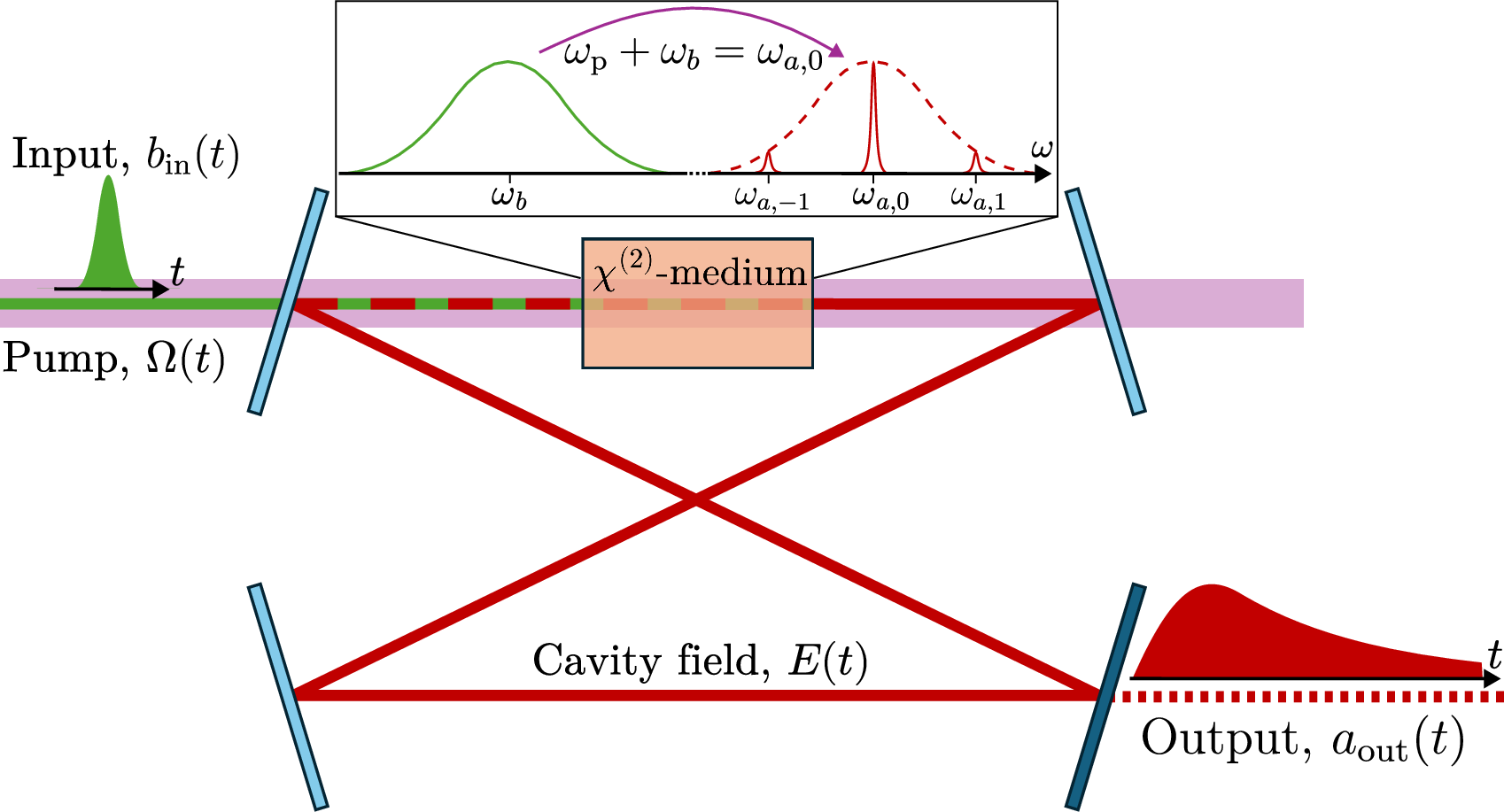}
    \caption{Illustration of the single-photon stretcher. A short input photon ($\omega_b$, green) and a classical pump ($\omega_\mathrm{p}$, purple), to which the cavity mirrors are transparent, undergo frequency conversion in the $\chi^{(2)}$-medium such that the input is converted to cavity mode at a different frequency ($\omega_{a,0}$, red). The cavity field is completely reflected by the three light-blue mirrors but is slowly leaking out of the dark-blue mirror, resulting in an effective stretching of the photon.
    The inset shows the process in the frequency domain where the broadband input is frequency-converted onto the narrowband cavity modes, and sidebands with $\omega_{a,\pm 1} = \omega_{a,0} \pm \FSR$.}
    \label{fig:intro}
\end{figure}

In this work, we introduce a solution based on perfect photon capture in a high-finesse cavity that simultaneously solves the frequency mismatch and bridges large bandwidth gaps, resulting in photon pulses compatible with typical hardware for quantum networks.
The approach relies on using cavity mirrors that are transparent to the input light, and controlled intra-cavity frequency conversion in a $\chi^{(2)}$-medium translating the input light to a frequency for which the cavity mirrors are highly reflecting.
This is illustrated in \cref{fig:intro}.
Once converted to the cavity frequency, the photon is stored inside the cavity on a time-scale set by the cavity decay rate, which can be much longer than the incoming pulse, resulting in an effective temporal stretching.
Related perfect photon capture techniques has previously been explored through recycling cavity reflections~\cite{wong_recycling_2025} and dynamical cavity coupling~\cite{krastanov_controlled-phase_2022}.
In contrast, our approach builds directly on established frequency‑conversion~\cite{murakami_low-noise_2025}, squeezed vacuum~\cite{shajilal_126_2022}, and photon‑generation techniques~\cite{wolfgramm_atom-resonant_2011,fekete_ultranarrow-band_2013,rielander_cavity_2016}, making it a natural extension of existing experimental platforms.

Optical cavities are known for their long decay times and are characterized by the finesse $\F$, which is proportional to the number of optical cycles of a photon in the cavity.
Given a roundtrip time of the cavity $\Trt$, the cavity decay-time is $\Tout = \F \Trt / 2\pi$~\cite{suter_calculation_2014}.
For an input pulse of duration $\Tin$, we therefore define the stretching factor
 \begin{equation}\label{eq:stretch_factor}
    \W = \frac{\Tout}{\Tin} = \frac{\F}{2\pi} \frac{\Trt}{\Tin}\,.
\end{equation}
The stretching factor depends solely on the cavity finesse and the ratio of the cavity round‑trip time to the input pulse duration.
We can thus achieve very high stretching factors if a short pulse is frequency converted in a high finesse cavity.
However, the finesse is ultimately limited by internal losses, while the ratio $\Trt / \Tin$ is constrained by both the physical cavity length and the onset of coupling to higher‑order longitudinal modes.
Below we will investigate how these factors limit the achievable stretching factor.

To achieve a large stretching factor, it is desirable to work with short pulses compared to the cavity round-trip $\Tin \ll \Trt$.
However, short pulses will be able to excite multiple longtudinal cavity modes:
after frequency conversion, a too short pulse would propagate freely around the cavity and only leaking out when it reaches the output mirror, producing periodic bursts in the output field, corresponding to the higher-order frequency components illustrated in the inset of \cref{fig:intro}.
For a smooth slowly varying output pulse we are thus limited to the regime $\Tin \gtrsim \Trt$, so that the stretching factor is predominantly limited by the achievable finesse $\F$.

To investigate the regime $\Tin \sim \Trt$, we further introduce the filtered stretching efficiency
\begin{equation}
    N_\mathrm{FSR} = \int_{-\frac{\FSR}{2}}^\frac{\FSR}{2}\d\omega\, |a_\mathrm{out}(\omega)|^2\,,
\end{equation}
where $a_\mathrm{out}(\omega)$ is the Fourier transform of the output field and $\FSR= \frac{2\pi}{\Trt}$ is the cavity free-spectral-range.
This quantity expresses the efficiency of converting an incoming photon into a frequency band of width $\FSR$, i.e. removing the contribution from cavity sidebands.
In the following sections, we build a theoretical model for the output field, from which we determine the optimal pump‑shape required for perfect photon capture.
Finally, we investigate the trade‑off between stretching factor and efficiency, and examine how internal losses limit the performance.

\section{Mathematical model}\label{sec:model}
We model the combined $\chi^{(2)}$-medium and cavity by super-positioning the frequency converted field over the number of cycles in the cavity.
This approach provides a completely general framework for a multi-mode cavity that is only limited by the approximations we impose on the frequency-conversion.

First, we describe the evolution of the field propagating through the non-linear crystal which couple the incident photon into the cavity.
This is governed by the well-known frequency conversion equations for a $\chi^{(2)}$-medium~\cite{reddy_temporal_2013}
\begin{subequations}
\begin{align}
    \left(\partial_z + \beta \partial_t\right) a(z,t) &= i\chi_2 \Omega(t - \beta z) b(z,t)\,,\\
    \left(\partial_z + \beta \partial_t\right) b(z,t) &= i\chi_2 \Omega^*(t - \beta z) a(z,t)\,.
\end{align}
\end{subequations}
Here we consider $a$ to be the field that couples to the cavity and $b$ the field to which the cavity mirrors are transparent.
$\chi_2$ is the nonlinear interaction parameter and $\Omega$ is the classical pump field.
We assume all fields to have identical inverse velocity $\beta$, and $t$ and $z$ denote, respectively, the time and the spatial coordinate along the propagation direction.
All fields are linear in the quantum operators, so we can describe the fields as complex numbers and leave out the operator notation~\cite{gorshkov_universal_2007}.
The solution can be expressed as
\begin{equation}
    a(t) = G_{aa}(t-\beta\ell) a_\mathrm{in}(t-\beta\ell) + G_{ab}(t-\beta\ell)b_\mathrm{in}(t-\beta\ell)\,,
\end{equation}
where $a_\mathrm{in}$ and $b_\mathrm{in}$ are the input fields and $\ell$ is the length of the nonlinear medium.
The solution for $b(t)$ is found by replacing $a\leftrightarrow b$.
The transfer functions describing the frequency conversion are given by
\begin{subequations}
\begin{align}
    G_{aa}(t) &= G_{bb}(t) = \cos(\chi_2\Omega(t) \ell)\,,\\
    G_{ab}(t) &= G_{ba}(t) = i\sin(\chi_2\Omega(t) \ell)\,.
\end{align}
\end{subequations}

We can now find the field inside the cavity, $E$, from the solution of the frequency conversion equations.
Summing up the electric field over the number of cycles through the cavity, we recursively find the cavity field as
\begin{subequations}
\begin{align}
    E(t) &= \sum_n E_n(t)\,,\\
    E_{n+1}(t) &= r G_{aa}(t-\beta\ell)E_{n}(t-\Trt)\,,\\ 
    E_0(t) &= G_{ab}(t-\beta\ell)b_\mathrm{in}(t-\beta\ell)\,,
\end{align}
\end{subequations}
where $r$ is the reflection coefficient of the output mirror.
The output field is given by
\begin{equation}\label{eq:output_a}
    a_\mathrm{out}(t) = \tilde{t}E(t-\beta_0L')\,,
\end{equation}
where $\tilde{t}$ is the transmission coefficient of the output mirror, $\beta_0$ is the inverse speed of light in air, and $L'$ is the distance from the nonlinear medium to the output mirror.
Similarly, we find the output of the field to which the cavity is transparent
\begin{equation}\label{eq:output_b}
\begin{aligned}
    b_\mathrm{out}(t) =& G_{bb}(t-\beta\ell)b_\mathrm{in}(t-\beta\ell)\\
    &+ r G_{ba}(t-\beta\ell)E(t - \Trt )\,.
\end{aligned}
\end{equation}

\begin{figure*}
    \centering
    \includegraphics[width=\textwidth]{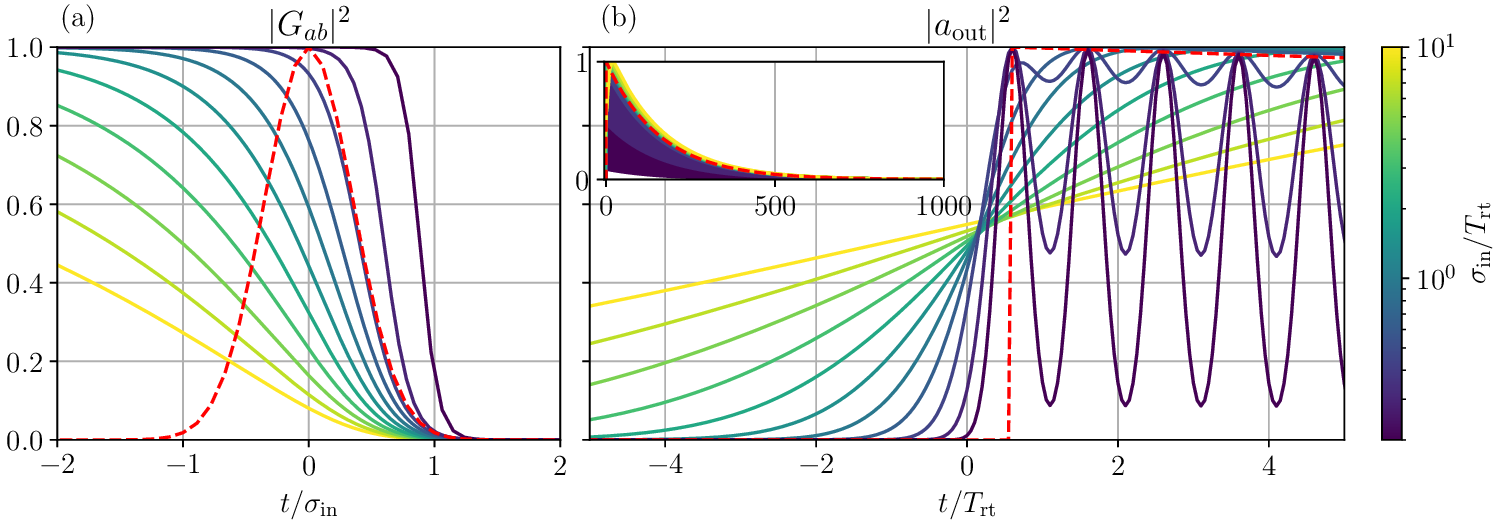}
    \caption{(a) The optimal transfer function $G_{ab}$ with the red dashed line being the input pulse. (b) The output intensity for short times normalized to its maximum. The inset shows the same, but on an extended timescale. The red dashed line indicates the single-mode approximation of the output. The color variation of the graphs in both (a) and (b) indicates the temporal width of the input, varying between $\Tin/\Trt = 0.1$ to $\Tin/\Trt = 10$ as given by the colorbar. We have assumed a cavity finesse of $\F = 1000$.}
    \label{fig:shapes}
\end{figure*}

We want to achieve perfect photon capture where the incoming photon is transferred to the cavity field with unit efficiency.
This requires the $b$-mode to be completely absorbed, meaning $b_\mathrm{out}(t)=0$.
Given this condition we solve for $\Omega$ in \cref{eq:output_b}, and find the optimal pump shape
\begin{equation}\label{eq:ideal_pump_shape}
    \Omega(t) = \frac{1}{\chi_2 \ell}\tan^{-1}\left(\frac{ib_\mathrm{in}(t)}{r E(t - \beta_0(L - \ell))}\right)\,.
\end{equation}
Here, the argument of $\tan^{-1}$ is real if the incoming field is real, since $E = i |E|$, and is infinite at $t=-\infty$, and 0 at $t=\infty$, resulting in $\tan^{-1}$ smoothly transitioning from $\frac{\pi}{2}$ to 0 over the course of the incoming pulse.

In the regime of very short pulses $\Tin \ll \Trt$, this behavior describes the pump initially being constant with $\Omega = \frac{\pi}{2\chi_2\ell}$ to achieve a perfect frequency conversion of the input.
After this, the pump is turned off as soon as the photon-pulse has passed through the non-linear medium, to avoid converting it back to the incoming frequency when the photon returns after one round trip.
For longer pulses $\Tin \gtrsim \Trt$, the shape of the pump field (\ref{eq:ideal_pump_shape}) ensures that the conversion of the incoming field and the field from the previous round trip interferes to achieve a perfect conversion from the incoming frequency to the cavity mode.

To find the exact shape of the pump in the regime $\Tin \gtrsim \Trt$, it is necessary to solve \cref{eq:ideal_pump_shape} iteratively.
As an initial guess of the pump shape, we assume that the input perfectly couples to the cavity, without cavity loss on the time-scale of the input such that $E \propto i\sqrt{\int_{-\infty}^td\tau |b_\mathrm{in}(\tau)|^2}$.
From this pump shape, we compute a new $E$, which is used to find a new pump shape.
Repeating this until $E$ has sufficiently converged, achieves the ideal shape of the pump $\Omega$.

\section{Results}\label{sec:results}
To investigate the performance of the photon stretcher, we will consider a Gaussian input pulse of the form
\begin{equation}
    b_\mathrm{in}(t) = \left(\frac{1}{2\pi\Tin^2}\right)^{\frac{1}{4}} \exp\left[-\frac{t^2}{4\Tin^2}\right]\,,
\end{equation}
where $\Tin$ is the input width defined as the standard deviation of the intensity.

In \cref{fig:shapes}(a), the optimal transfer functions $G_{ab}$ are shown for different choices of input width normalized to roundtrip time, and in \cref{fig:shapes}(b), the corresponding output fields are displayed.
When $\Tin \ll \Trt$, we observe the expected behavior: the transfer function behaves nearly as a step-function, with perfect frequency-conversion while the photon-pulse is incident and is then rapidly turned to zero before the frequency converted photon performs a round trip, as seen in \cref{fig:shapes}(a).
The fact that the short incoming pulse is converted to the cavity frequency as a short pulse results in a traveling wave packet inside the cavity that populates multiple cavity modes, leading to a highly oscillatory output field, as visible in \cref{fig:shapes}(b).

\begin{figure}[t]
    \centering
    \includegraphics[width=\linewidth]{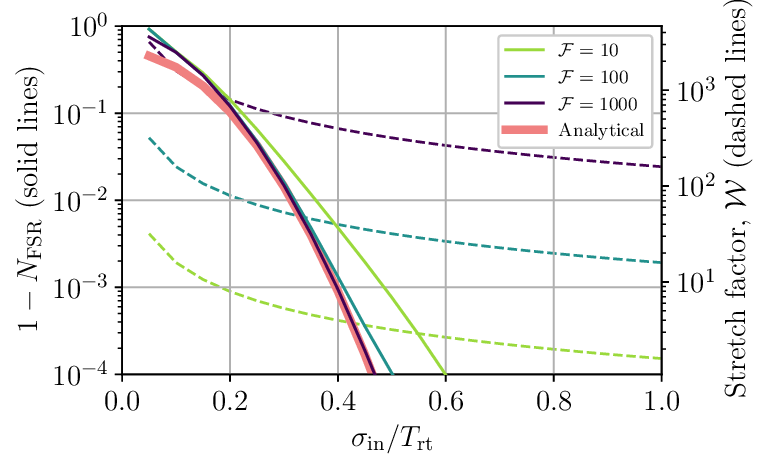}
    \caption{The solid lines (left axis) shows the inefficiency of the coupling into the fundamental mode of the cavity as a function of input width normalized to the cavity roundtrip time for three different cavity finesses indicated by the color. These are compared to the analytical approximation $\F \rightarrow \infty$ illustrated by the red line. The dashed lines (right axis) shows the corresponding stretching factor.}
    \label{fig:efficiency}
\end{figure}

A more smooth output field is obtained for $\Tin \gg \Trt$.
In this case the input wraps around the cavity several times while the pulse is incident, so that a less efficient frequency conversion is necessary to avoid converting a photon from $a$ back to $b$.
This can be seen as the gradual decrease of $G_{ab}$ in \cref{fig:shapes}(a) as the pulse gets longer.
As a result we observe a smooth behavior of the output field in \cref{fig:shapes}(b) due to the input strictly coupling to the fundamental mode.
The inset in \cref{fig:shapes}(a) shows the output field on a longer time scale, which better visualizes how the input field with a width $\Tin = 0.1\text{-}10\Trt$, is extended to a width on the order of $\sim 100\Trt$, determined by the finesse $\F/2\pi \approx 150$.

In \cref{fig:efficiency}, we show the stretching inefficiency $1-N_\mathrm{FSR}$ for three choices of the finesse.
For $\Tin \ll \Trt$ the efficiency is small due to coupling to higher-order cavity modes.
As $\Tin / \Trt$ increases, the coupling to the fundamental mode is better, giving rise to a smaller stretching inefficiency.
At the same time, the stretch factor decreases, simply because the pulse is longer compared to the fixed decay time of the cavity. 
There is a slight difference in efficiency depending on finesse; this is due to the spectral tails of the fundamental mode that extend beyond the square filter.
In the Supplementary Material, we develop an analytical model describing the regime $\F \rightarrow\infty$.
Including the first side band on each side of the fundamental mode, the stretching efficiency is approximated as $N_\mathrm{FSR} \approx 1 - \frac{1}{2}\exp\left[-\left(2\pi\Tin/\Trt\right)^2\right]$, which is very in good agreement with the numerical results.
Due to the rapid decrease of the Gaussian, already at $\Tin > 0.5\Trt$, the stretching inefficiency is less than $10^{-3}$.
Thus, we can achieve a stretching factor $\W \sim \F/\pi$ while having near-perfect coupling to the fundamental cavity mode.

\begin{figure}[b]
    \centering
    \includegraphics[width=1\linewidth]{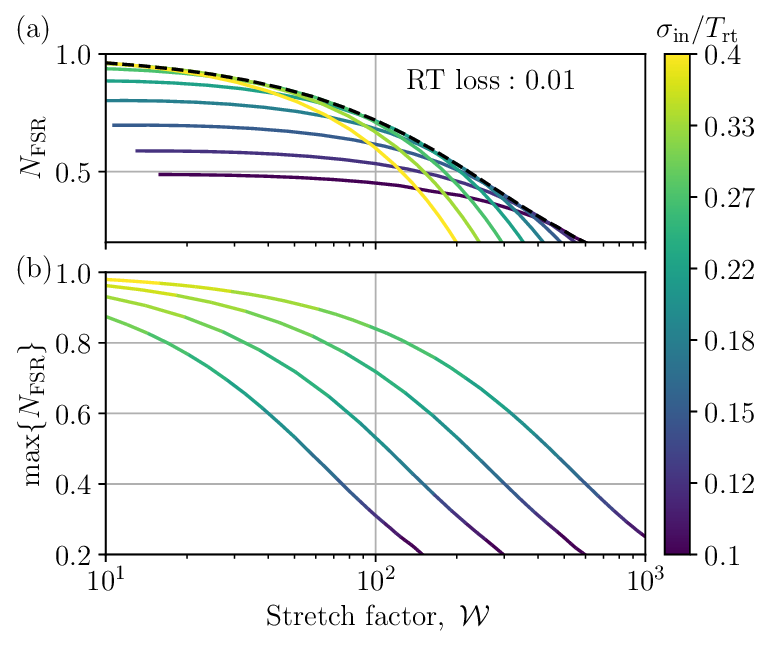}
    \caption{(a) Stretching efficiency for a roundtrip (RT) loss of 0.01 as a function of the stretch factor (obtained by varying the cavity finesse) for different choices of $\Tin/\Trt$ as indicated by the colorbar. Each colorbar tickmark corresponds to a curve. The black dashed line shows the maximum $N_\mathrm{FSR}$ value of the graphs, corresponding to the maximum achievable efficiency for a system with the given RT loss and stretching factor. (b) The maximum achievable $N_\mathrm{FSR}$ for a given stretching factor for four choices of RT loss: counting from above $l = 0.005$, 0.01, 0.02, 0.04. The colorgrading indicates the value of $\Tin/\Trt$.}
    \label{fig:loss}
\end{figure}

\subsection{Roundtrip loss}
Above we have found almost perfect stretching set by the cavity finesse.
In practice, however, the finesse is typically limited by losses.
To understand how losses degrade the performance of the photon stretcher, we introduce a roundtrip loss to our system by defining the reflection coefficient as
\begin{equation}
    r = \tilde{r}\sqrt{1-l}
\end{equation}
where $\tilde{r}$ is the reflection coefficient of the output mirror, and $l$ is the internal roundtrip loss probability.
Note that the transmission coefficient in \cref{eq:output_a} is now given by $\tilde{t} = \sqrt{1-\tilde{r}^2}$.
The internal roundtrip loss can potentially describe losses due to imperfect mirrors, reflections at the boundary of the nonlinear medium, or propagation losses during frequency conversion.

To understand how the photon stretching affect the efficiency in the presence of internal loss, we compute $N_\mathrm{FSR}$ as a function of the stretching factor (by varying the finesse) for several choices of $\Tin/\Trt$, for a fixed round-trip loss as shown in \cref{fig:loss}(a).
The efficiency $N_\mathrm{FSR}$ decreases with stretching factor due to an increased number of roundtrips resulting in a larger effective loss, and is ultimately limited by
\begin{equation}
    \W_\mathrm{max} = \frac{\F_\mathrm{loss}}{2\pi}\frac{\Trt}{\Tin}
\end{equation}
where $\F_\mathrm{loss} = \frac{\pi}{|\ln(\sqrt{1-l})|}$ is the maximal finesse achievable given the loss.
Evidently, the maximum stretching factor will therefore increase as $\Tin/\Trt$ decreases, as shown by the different colored lines in \cref{fig:loss}(a).

As is evident from \cref{fig:loss}(a) there is a trade off between stretching factor and efficiency.
To achieve the optimal efficiency for a given stretching factor we thus need to choose the value of $\Tin/\Trt$ with highest efficiency $N_\mathrm{FSR}$.
This is given by the black dashed line in \cref{fig:loss}(a).
Physically, this corresponds to tuning the cavity length and output mirror transmission for a given input width to achieve the maximum stretching efficiency.

In \cref{fig:loss}(b) we present this optimum for four different values of round trip loss.
Each line is color graded according to the corresponding $\Tin/\Trt$ that gives rise to this point on the graph.
From these graphs, the required cavity design parameters to achieve a certain efficiency given a roundtrip loss can be deduced.
As an example, consider a system with a loss of 0.01, and a desired stretching factor of 100.
An efficiency of 0.7 is achieved for $\Tin/\Trt\sim 0.27$.
From \cref{eq:stretch_factor}, the cavity finesse has to be $\sim 170$.
For an input width of $\Tin = 1\,\mathrm{ns}$, corresponding to a quantum dot single photon source~\cite{uppu_scalable_2020}, we need a roundtrip time of  $\Trt \approx 3.8\,\mathrm{ns}$ corresponding to an effective cavity length of $L_\mathrm{eff} = c\Trt \approx 1.1\,\mathrm{m}$.
These values are comparable to Ref.~\cite{rielander_cavity_2016}, that report on an almost identical setup with cavity length of 0.7\,m and finesse of $\sim 145$, for single-photon generation.
This underscores the experimental feasibility of our proposed approach.

\section{Conclusion}\label{sec:conclusion}
In conclusion, we have introduced and modeled a system that facilitates a coherent stretching of photon pulses.
The stretching is enabled by a $\chi^{(2)}$-medium that ensures perfect coupling to a high-finesse cavity, from which the slow leakage results in an effective stretching only limited by the cavity finesse and internal losses.
The proposed setup can serve as a versatile tool for interfacing different physical systems by combining frequency conversion and bandwidth matching.
Specifically we have shown that it is possible convert a broadband quantum dot single photon to match a narrowband solid state memory.
Such an interface could be useful for building quantum repeaters that are essential building blocks for large scale quantum communication~\cite{azuma_quantum_2023}.

\section*{Acknowledgments}
This work was funded by the European Union's Horizon Europe research and innovation programme under grant agreement No. 101102140 – QIA Phase 1.
We acknowledge funding from the Danish National Research Foundation (Center of Excellence “Hy-Q,” Grant No. DNRF139) and the Novo Nordisk Foundation (Challenge project “SolidQ”, Grant No. NNF20OC0060019).

Funded by the European Union. Views and opinions expressed are however those of the author(s) only and do not necessarily reflect those of the European Union or European Commission. Neither the European Union nor the granting authority can be held responsible for them.


\bibliography{references}

@misc{mohan_deterministic_2026,
	title = {Deterministic photon waveform adaptation for quantum connectivity},
	url = {http://arxiv.org/abs/2609.26889},
	doi = {10.48550/arXiv.2609.26889},
	urldate = {2026-10-05},
	publisher = {arXiv},
	author = {Mohan, Jeffrey and Berroir, Jérémy and Borselli, Filippo and Libault, David and Loubar, Ferhat and Müller, Kilian and Rowland, Jed and Hoffet, Félix and Diamanti, Eleni and Darras, Tom and Mazzoni, Tommaso and Laurat, Julien},
	month = sep,
	year = {2026},
	note = {arXiv:2609.26889 [quant-ph]},
}

@article{krastanov_controlled-phase_2022,
	title = {Controlled-phase gate by dynamic coupling of photons to a two-level emitter},
	volume = {8},
	copyright = {2022 The Author(s)},
	issn = {2056-6387},
	url = {https://www.nature.com/articles/s41534-022-00604-5},
	doi = {10.1038/s41534-022-00604-5},
	language = {en},
	number = {1},
	urldate = {2026-09-14},
	journal = {npj Quantum Information},
	publisher = {Nature Publishing Group},
	author = {Krastanov, Stefan and Jacobs, Kurt and Gilbert, Gerald and Englund, Dirk R. and Heuck, Mikkel},
	month = sep,
	year = {2022},
	pages = {103},
}

@article{afzelius_multimode_2009,
	title = {Multimode quantum memory based on atomic frequency combs},
	volume = {79},
	url = {https://link.aps.org/doi/10.1103/PhysRevA.79.052329},
	doi = {10.1103/PhysRevA.79.052329},
	number = {5},
	urldate = {2026-09-14},
	journal = {Physical Review A},
	publisher = {American Physical Society},
	author = {Afzelius, Mikael and Simon, Christoph and de Riedmatten, Hugues and Gisin, Nicolas},
	month = may,
	year = {2009},
	pages = {052329},
}

@article{hanni_heralded_2025,
	title = {Heralded {Entanglement} of {On}-{Demand} {Spin}-{Wave} {Solid}-{State} {Quantum} {Memories} for {Multiplexed} {Quantum} {Network} {Links}},
	volume = {15},
	url = {https://link.aps.org/doi/10.1103/wvv1-6lg8},
	doi = {10.1103/wvv1-6lg8},
	number = {4},
	urldate = {2026-09-14},
	journal = {Physical Review X},
	publisher = {American Physical Society},
	author = {Hänni, Jonathan and Rodríguez-Moldes, Alberto E. and Appas, Félicien and Wengerowsky, Soeren and Lago-Rivera, Dario and Teller, Markus and Grandi, Samuele and de Riedmatten, Hugues},
	month = oct,
	year = {2025},
	pages = {041003},
}

@article{azuma_quantum_2023,
	title = {Quantum repeaters: {From} quantum networks to the quantum internet},
	volume = {95},
	issn = {0034-6861, 1539-0756},
	shorttitle = {Quantum repeaters},
	url = {https://link.aps.org/doi/10.1103/RevModPhys.95.045006},
	doi = {10.1103/RevModPhys.95.045006},
	language = {en},
	number = {4},
	urldate = {2026-06-16},
	journal = {Reviews of Modern Physics},
	author = {Azuma, Koji and Economou, Sophia E. and Elkouss, David and Hilaire, Paul and Jiang, Liang and Lo, Hoi-Kwong and Tzitrin, Ilan},
	month = dec,
	year = {2023},
	pages = {045006},
}

@article{teller_solid-state_2025,
	title = {A solid-state temporally multiplexed quantum memory array at the single-photon level},
	volume = {11},
	copyright = {2025 The Author(s)},
	issn = {2056-6387},
	url = {https://www.nature.com/articles/s41534-025-01042-9},
	doi = {10.1038/s41534-025-01042-9},
	language = {en},
	number = {1},
	urldate = {2026-06-15},
	journal = {npj Quantum Information},
	publisher = {Nature Publishing Group},
	author = {Teller, Markus and Plascencia, Susana and Sastre Jachimska, Cristina and Grandi, Samuele and de Riedmatten, Hugues},
	month = jun,
	year = {2025},
	pages = {92},
}

@article{gorshkov_universal_2007,
	title = {Universal {Approach} to {Optimal} {Photon} {Storage} in {Atomic} {Media}},
	volume = {98},
	url = {https://link.aps.org/doi/10.1103/PhysRevLett.98.123601},
	doi = {10.1103/PhysRevLett.98.123601},
	number = {12},
	urldate = {2025-08-05},
	journal = {Physical Review Letters},
	publisher = {American Physical Society},
	author = {Gorshkov, Alexey V. and André, Axel and Fleischhauer, Michael and Sørensen, Anders S. and Lukin, Mikhail D.},
	month = mar,
	year = {2007},
	pages = {123601},
}

@article{vernaz-gris_highly-efficient_2018,
	title = {Highly-efficient quantum memory for polarization qubits in a spatially-multiplexed cold atomic ensemble},
	volume = {9},
	copyright = {2018 The Author(s)},
	issn = {2041-1723},
	url = {https://www.nature.com/articles/s41467-017-02775-8},
	doi = {10.1038/s41467-017-02775-8},
	language = {en},
	number = {1},
	urldate = {2026-06-15},
	journal = {Nature Communications},
	publisher = {Nature Publishing Group},
	author = {Vernaz-Gris, Pierre and Huang, Kun and Cao, Mingtao and Sheremet, Alexandra S. and Laurat, Julien},
	month = jan,
	year = {2018},
	pages = {363},
}

@article{kuhlmann_charge_2013,
	title = {Charge noise and spin noise in a semiconductor quantum device},
	volume = {9},
	copyright = {2013 Springer Nature Limited},
	issn = {1745-2481},
	url = {https://www.nature.com/articles/nphys2688},
	doi = {10.1038/nphys2688},
	language = {en},
	number = {9},
	urldate = {2026-06-15},
	journal = {Nature Physics},
	publisher = {Nature Publishing Group},
	author = {Kuhlmann, Andreas V. and Houel, Julien and Ludwig, Arne and Greuter, Lukas and Reuter, Dirk and Wieck, Andreas D. and Poggio, Martino and Warburton, Richard J.},
	month = sep,
	year = {2013},
	pages = {570--575},
}

@article{cho_highly_2016,
	title = {Highly efficient optical quantum memory with long coherence time in cold atoms},
	volume = {3},
	copyright = {https://doi.org/10.1364/OA\_License\_v1\#VOR-OA},
	issn = {2334-2536},
	url = {https://opg.optica.org/abstract.cfm?URI=optica-3-1-100},
	doi = {10.1364/OPTICA.3.000100},
	language = {en},
	number = {1},
	urldate = {2026-06-15},
	journal = {Optica},
	author = {Cho, Y.-W. and Campbell, G. T. and Everett, J. L. and Bernu, J. and Higginbottom, D. B. and Cao, M. T. and Geng, J. and Robins, N. P. and Lam, P. K. and Buchler, B. C.},
	month = jan,
	year = {2016},
	pages = {100},
}

@article{allgaier_highly_2017,
	title = {Highly efficient frequency conversion with bandwidth compression of quantum light},
	volume = {8},
	copyright = {2017 The Author(s)},
	issn = {2041-1723},
	url = {https://www.nature.com/articles/ncomms14288},
	doi = {10.1038/ncomms14288},
	language = {en},
	number = {1},
	urldate = {2026-05-06},
	journal = {Nature Communications},
	publisher = {Nature Publishing Group},
	author = {Allgaier, Markus and Ansari, Vahid and Sansoni, Linda and Eigner, Christof and Quiring, Viktor and Ricken, Raimund and Harder, Georg and Brecht, Benjamin and Silberhorn, Christine},
	month = jan,
	year = {2017},
	pages = {14288},
}

@article{engelhardt_optimal_2022,
	title = {Optimal {Broadband} {Frequency} {Conversion} via a {Magnetomechanical} {Transducer}},
	volume = {18},
	doi = {10.1103/PhysRevApplied.18.044059},
	number = {4},
	journal = {Physical Review Applied},
	author = {Engelhardt, F.},
	year = {2022},
}

@article{anetsberger_ultralow-dissipation_2008,
	title = {Ultralow-dissipation optomechanical resonators on a chip},
	volume = {2},
	copyright = {2008 Springer Nature Limited},
	issn = {1749-4893},
	url = {https://www.nature.com/articles/nphoton.2008.199},
	doi = {10.1038/nphoton.2008.199},
	language = {en},
	number = {10},
	urldate = {2026-05-06},
	journal = {Nature Photonics},
	publisher = {Nature Publishing Group},
	author = {Anetsberger, G. and Rivière, R. and Schliesser, A. and Arcizet, O. and Kippenberg, T. J.},
	month = oct,
	year = {2008},
	pages = {627--633},
}

@article{lauk_perspectives_2020,
	title = {Perspectives on quantum transduction},
	volume = {5},
	issn = {2058-9565},
	url = {https://doi.org/10.1088/2058-9565/ab788a},
	doi = {10.1088/2058-9565/ab788a},
	language = {en},
	number = {2},
	urldate = {2026-04-27},
	journal = {Quantum Science and Technology},
	publisher = {IOP Publishing},
	author = {Lauk, Nikolai and Sinclair, Neil and Barzanjeh, Shabir and Covey, Jacob P and Saffman, Mark and Spiropulu, Maria and Simon, Christoph},
	month = mar,
	year = {2020},
	pages = {020501},
}

@article{cirac_quantum_1997,
	title = {Quantum {State} {Transfer} and {Entanglement} {Distribution} among {Distant} {Nodes} in a {Quantum} {Network}},
	volume = {78},
	doi = {10.1103/PhysRevLett.78.3221},
	number = {16},
	journal = {Physical Review Letters},
	author = {Cirac, J. I.},
	year = {1997},
	pages = {3221--3224},
}

@article{murakami_low-noise_2025,
	title = {Low-noise quantum frequency conversion with cavity enhancement of the converted mode},
	volume = {3},
	issn = {2837-6714},
	url = {https://opg.optica.org/abstract.cfm?URI=opticaq-3-1-55},
	doi = {10.1364/OPTICAQ.542350},
	language = {en},
	number = {1},
	urldate = {2025-11-07},
	journal = {Optica Quantum},
	author = {Murakami, Shoichi and Kobayashi, Toshiki and Miki, Shigehito and Terai, Hirotaka and Kodama, Tsuyoshi and Sawaya, Tsuneaki and Ohtomo, Akihiko and Shimoi, Hideki and Yamamoto, Takashi and Ikuta, Rikizo},
	month = feb,
	year = {2025},
	pages = {55},
}

@article{kimble_quantum_2008,
	title = {The quantum internet},
	volume = {453},
	copyright = {2008 Springer Nature Limited},
	issn = {1476-4687},
	url = {https://www.nature.com/articles/nature07127},
	doi = {10.1038/nature07127},
	language = {en},
	number = {7198},
	urldate = {2026-04-27},
	journal = {Nature},
	publisher = {Nature Publishing Group},
	author = {Kimble, H. J.},
	month = jun,
	year = {2008},
	pages = {1023--1030},
}

@article{wong_recycling_2025,
	title = {Recycling reflections for perfect photon capture},
	volume = {112},
	issn = {2469-9926, 2469-9934},
	url = {https://link.aps.org/doi/10.1103/ysnl-8krl},
	doi = {10.1103/ysnl-8krl},
	language = {en},
	number = {3},
	urldate = {2025-11-07},
	journal = {Physical Review A},
	author = {Wong, Yat and Jiang, Liang},
	month = sep,
	year = {2025},
	pages = {032412},
}

@article{fekete_ultranarrow-band_2013,
	title = {Ultranarrow-{Band} {Photon}-{Pair} {Source} {Compatible} with {Solid} {State} {Quantum} {Memories} and {Telecommunication} {Networks}},
	volume = {110},
	copyright = {http://link.aps.org/licenses/aps-default-license},
	issn = {0031-9007, 1079-7114},
	url = {https://link.aps.org/doi/10.1103/PhysRevLett.110.220502},
	doi = {10.1103/PhysRevLett.110.220502},
	language = {en},
	number = {22},
	urldate = {2026-01-30},
	journal = {Physical Review Letters},
	author = {Fekete, Julia and Rieländer, Daniel and Cristiani, Matteo and De Riedmatten, Hugues},
	month = may,
	year = {2013},
	pages = {220502},
}

@article{wolfgramm_atom-resonant_2011,
	title = {Atom-{Resonant} {Heralded} {Single} {Photons} by {Interaction}-{Free} {Measurement}},
	volume = {106},
	copyright = {http://link.aps.org/licenses/aps-default-license},
	issn = {0031-9007, 1079-7114},
	url = {https://link.aps.org/doi/10.1103/PhysRevLett.106.053602},
	doi = {10.1103/PhysRevLett.106.053602},
	language = {en},
	number = {5},
	urldate = {2026-01-30},
	journal = {Physical Review Letters},
	author = {Wolfgramm, Florian and De Icaza Astiz, Yannick A. and Beduini, Federica A. and Cerè, Alessandro and Mitchell, Morgan W.},
	month = feb,
	year = {2011},
	pages = {053602},
}

@article{bussieres_prospective_2013,
	title = {Prospective applications of optical quantum memories},
	volume = {60},
	issn = {0950-0340},
	url = {https://doi.org/10.1080/09500340.2013.856482},
	doi = {10.1080/09500340.2013.856482},
	number = {18},
	urldate = {2026-01-30},
	journal = {Journal of Modern Optics},
	publisher = {Taylor \& Francis},
	author = {Bussières, Félix and Sangouard, Nicolas and Afzelius, Mikael and de Riedmatten, Hugues and Simon, Christoph and Tittel, Wolfgang},
	month = oct,
	year = {2013},
	note = {\_eprint: https://doi.org/10.1080/09500340.2013.856482},
	pages = {1519--1537},
}

@article{afzelius_quantum_2015,
	title = {Quantum memory for photons},
	volume = {68},
	issn = {0031-9228, 1945-0699},
	url = {https://physicstoday.aip.org/features/quantum-memory-for-photons},
	doi = {10.1063/PT.3.3021},
	language = {en},
	number = {12},
	urldate = {2026-01-30},
	journal = {Physics Today},
	author = {Afzelius, Mikael and Gisin, Nicolas and De Riedmatten, Hugues},
	month = dec,
	year = {2015},
	pages = {42--47},
}

@article{shajilal_126_2022,
	title = {12.6 {dB} squeezed light at 1550 nm from a bow-tie cavity for long-term high duty cycle operation},
	volume = {30},
	copyright = {© 2022 Optica Publishing Group},
	issn = {1094-4087},
	url = {https://opg.optica.org/oe/abstract.cfm?uri=oe-30-21-37213},
	doi = {10.1364/OE.465521},
	language = {EN},
	number = {21},
	urldate = {2026-01-30},
	journal = {Optics Express},
	publisher = {Optica Publishing Group},
	author = {Shajilal, Biveen and Thearle, Oliver and Tranter, Aaron and Lu, Yuerui and Huntington, Elanor and Assad, Syed and Lam, Ping Koy and Janousek, Jiri},
	month = oct,
	year = {2022},
	pages = {37213--37223},
}

@article{rielander_cavity_2016,
	title = {Cavity enhanced telecom heralded single photons for spin-wave solid state quantum memories},
	volume = {18},
	issn = {1367-2630},
	url = {https://doi.org/10.1088/1367-2630/aa4f38},
	doi = {10.1088/1367-2630/aa4f38},
	language = {en},
	number = {12},
	urldate = {2025-11-14},
	journal = {New Journal of Physics},
	publisher = {IOP Publishing},
	author = {Rieländer, Daniel and Lenhard, Andreas and Mazzera, Margherita and Riedmatten, Hugues de},
	month = dec,
	year = {2016},
	pages = {123013},
}

@article{reddy_temporal_2013,
	title = {Temporal mode selectivity by frequency conversion in second-order nonlinear optical waveguides},
	volume = {21},
	copyright = {https://doi.org/10.1364/OA\_License\_v1\#VOR-OA},
	issn = {1094-4087},
	url = {https://opg.optica.org/oe/abstract.cfm?uri=oe-21-11-13840},
	doi = {10.1364/OE.21.013840},
	language = {en},
	number = {11},
	urldate = {2025-08-12},
	journal = {Optics Express},
	author = {Reddy, D. V. and Raymer, M. G. and McKinstrie, C. J. and Mejling, L. and Rottwitt, K.},
	month = jun,
	year = {2013},
	pages = {13840},
}

@article{heshami_quantum_2016,
	title = {Quantum memories: emerging applications and recent advances},
	volume = {63},
	issn = {0950-0340},
	shorttitle = {Quantum memories},
	url = {https://doi.org/10.1080/09500340.2016.1148212},
	doi = {10.1080/09500340.2016.1148212},
	number = {20},
	urldate = {2026-01-20},
	journal = {Journal of Modern Optics},
	publisher = {Taylor \& Francis},
	author = {Heshami, Khabat and England, Duncan G. and Humphreys, Peter C. and Bustard, Philip J. and Acosta, Victor M. and Nunn, Joshua and Sussman, Benjamin J.},
	month = nov,
	year = {2016},
	note = {\_eprint: https://doi.org/10.1080/09500340.2016.1148212},
	pages = {2005--2028},
}

@article{divincenzo_physical_2000,
	title = {The {Physical} {Implementation} of {Quantum} {Computation}},
	volume = {48},
	issn = {1521-3978},
	doi = {10.1002/1521-3978(200009)48:9/11<771::AID-PROP771>3.0.CO;2-E},
	language = {en},
	number = {9-11},
	urldate = {2024-12-16},
	journal = {Fortschritte der Physik},
	author = {DiVincenzo, David P.},
	year = {2000},
	pages = {771--783},
}

@article{uppu_quantum-dot-based_2021,
	title = {Quantum-dot-based deterministic photon–emitter interfaces for scalable photonic quantum technology},
	volume = {16},
	copyright = {2021 Springer Nature Limited},
	issn = {1748-3395},
	url = {https://www.nature.com/articles/s41565-021-00965-6},
	doi = {10.1038/s41565-021-00965-6},
	language = {en},
	number = {12},
	urldate = {2026-01-20},
	journal = {Nature Nanotechnology},
	publisher = {Nature Publishing Group},
	author = {Uppu, Ravitej and Midolo, Leonardo and Zhou, Xiaoyan and Carolan, Jacques and Lodahl, Peter},
	month = dec,
	year = {2021},
	pages = {1308--1317},
}

@article{uppu_scalable_2020,
	title = {Scalable integrated single-photon source},
	volume = {6},
	url = {https://www.science.org/doi/10.1126/sciadv.abc8268},
	doi = {10.1126/sciadv.abc8268},
	number = {50},
	urldate = {2026-01-20},
	journal = {Science Advances},
	publisher = {American Association for the Advancement of Science},
	author = {Uppu, Ravitej and Pedersen, Freja T. and Wang, Ying and Olesen, Cecilie T. and Papon, Camille and Zhou, Xiaoyan and Midolo, Leonardo and Scholz, Sven and Wieck, Andreas D. and Ludwig, Arne and Lodahl, Peter},
	month = dec,
	year = {2020},
	pages = {eabc8268},
}

@article{rossi_measurement-based_2018,
	title = {Measurement-based quantum control of mechanical motion},
	volume = {563},
	copyright = {2018 Springer Nature Limited},
	issn = {1476-4687},
	url = {https://www.nature.com/articles/s41586-018-0643-8},
	doi = {10.1038/s41586-018-0643-8},
	language = {en},
	number = {7729},
	urldate = {2026-01-20},
	journal = {Nature},
	publisher = {Nature Publishing Group},
	author = {Rossi, Massimiliano and Mason, David and Chen, Junxin and Tsaturyan, Yeghishe and Schliesser, Albert},
	month = nov,
	year = {2018},
	pages = {53--58},
}

@article{suter_calculation_2014,
	title = {Calculation of the finesse of an ideal {Fabry}–{Perot} resonator},
	volume = {53},
	issn = {1559-128X, 2155-3165},
	url = {https://opg.optica.org/abstract.cfm?URI=ao-53-30-7004},
	doi = {10.1364/AO.53.007004},
	language = {en},
	number = {30},
	urldate = {2025-10-13},
	journal = {Applied Optics},
	author = {Suter, Martin and Dietiker, Peter},
	month = oct,
	year = {2014},
	pages = {7004},
}

\clearpage 
\onecolumngrid 

\begin{center}
\textbf{\large Supplementary Material for\\ "Temporal stretching of single-photon pulses by intra-cavity frequency conversion"}\\[1ex]
Mathias L. H. Korsgaard, Eva M. González-Ruiz, and Anders S. Sørensen\\
{\small \textit{Center for Hybrid Quantum Networks (Hy-Q), Niels Bohr Institute,\\ University of Copenhagen, Jagtvej 155A, 2200 Copenhagen N, Denmark}}
\end{center}

\setcounter{equation}{0} 
\renewcommand{\theequation}{S\arabic{equation}} 
\setcounter{figure}{0} 
\renewcommand{\thefigure}{S\arabic{figure}}

\renewcommand{\a}{\hat{a}}
\renewcommand{\b}{\hat{b}}
\renewcommand{\c}{\hat{c}}

Here we describe the efficiency of the single-photon stretcher detailed in the main text using an analytical model.
To do so, we consider the cavity to be described by a set of orthogonal modes $\{\a_j\}$ each representing a longitudinal mode of the cavity, corresponding to the sidebands shown in the inset of \cref{fig:intro}.
We will consider the limit $\F \rightarrow \infty$ where we can treat the modes independently, neglecting any interference effect in the emission from the cavity.
The modes interact with another cavity field described by $\b$ representing the input light, with a time-dependent interaction strength $\Omega(t)$ representing the classical pump.
The Hamiltonian governing this is
\begin{equation}
    \hat{H} = \hbar \omega_b \b^\dagger\b + \hbar\sum_j \omega_{a,j} \a_j^\dagger \a_j + \hbar \sum_j \left[ \Omega (t) \a_j^\dagger\b + \mathrm{h.c.}\right]
\end{equation}
where $\omega_{a,j} = \omega_{a,} + \Delta_{a,j}$ and $\Delta_{a,j} = j\FSR$.

To compare with the numerical results in the main text, we want to determine the filtered stretching efficiency.
In the regime $\F \rightarrow \infty$ this corresponds to the population of the fundamental mode at $t=\infty$ given by
\begin{equation}
    N_\mathrm{FSR} = \expval{\a_0^\dagger(t=\infty)\a_0(t=\infty)}\,.
\end{equation}

To model the system we will be using the input-output formalism, where we consider a single-photon input state
\begin{equation}
    \ket{\psi_\mathrm{in}} = \int_{-\infty}^\infty \d t\, \psi_\mathrm{in}(t) \b_\mathrm{in}^\dagger(t) \ket{0}
\end{equation}
The Heisenberg equations of motion in the input-output formalism are given by
\begin{equation}
\begin{aligned}
    \partial_t \a_j &= \frac{i}{\hbar}\left[\hat{H},\a_j\right] - \frac{\kappa_a}{2}\a_j \\
    \partial_t \b &= \frac{i}{\hbar}\left[\hat{H},\b\right] - \frac{\kappa_b}{2}\b + \sqrt{\kappa_{b}}\b_\mathrm{in}
\end{aligned}
\end{equation}
where $\kappa_a$ is the linewidth of the cavity and $\kappa_b$ is the linewidth of the virtual cavity mode $\b$.
Since we consider the regime $\F \rightarrow \infty$ we let $\kappa_a \rightarrow 0$ immediately, and later let $\kappa_b \rightarrow \infty$ so that the virtual cavity can represent the propagating input mode.
We use a rotating frame with $\b$ and $\b_\mathrm{in}$ rotating at a center frequency $\omega_b$, i.e. $\b \rightarrow \b \e^{-i\omega_b t}$.
Similarly we choose $\a_j$ to rotate at the frequency of the center frequency $\omega_{a,0}$, i.e. $\a_j \rightarrow \a_j \e^{-i\omega_{a,0}t}$.
Finally, we introduce the rotating frequency of the pump $\Omega \rightarrow \Omega \e^{-i\omega_\mathrm{p} t}$ and use energy conservation of the frequency conversion $\omega_\mathrm{p} + \omega_b - \omega_{a,0} = 0$ and the equations then become
\begin{equation}
\begin{aligned}
    \partial_t \a_j &= -i\Delta_{a,j}\a_j - i\Omega(t)\b  \\
    \partial_t \b &= - i\sum_j \Omega(t) \a_j - \frac{\kappa_b}{2}\b + \sqrt{\kappa_b}\b_\mathrm{in}
\end{aligned}
\end{equation}

We start by solving the equation for $\b$ using adiabatic elimination since we are interested in the limit $\kappa_b \rightarrow \infty$.
The solution is then
\begin{equation}
    \b(t) = \frac{2}{\kappa_b} \left(\sqrt{\kappa_b}\b_\mathrm{in}(t) - i \sum_j\Omega(t) \a_j(t)\right)
\end{equation}
Using this, we arrive at the final equation for the longitudinal modes of the cavity
\begin{equation}
    \partial_t \a_j = -\left(i\Delta_{a,j} + \frac{2}{\kappa_b}\Omega^2(t)\right)\a_j- i\frac{2}{\sqrt{\kappa_b}}\Omega(t) \b_\mathrm{in}(t) - \frac{2}{\kappa_b}\Omega^2(t) \sum_{k \neq j} \a_k(t)
\end{equation}
We can solve this equation perturbatively in the mode-coupling by introducing a perturbation parameter $\gamma$ on the last term and expressing the solution as $\a_j \approx \a_j^{(0)} + \gamma \a_j^{(1)}$.
Inserting this and collecting the terms in powers of $\gamma$ we end up with two equations
\begin{align}
    \partial_t \a_j^{(0)} &= -\left(i\Delta_{a,j} + \frac{2}{\kappa_b}\Omega^2(t)\right)\a_j^{(0)}- i\frac{2}{\sqrt{\kappa_b}}\Omega(t) \b_\mathrm{in}(t)\\
    \partial_t \a_j^{(1)} &=  - \frac{2}{\kappa_b}\Omega^2(t) \sum_{k \neq j} \a_k^{(0)}(t)
\end{align}
Solving these equations we can then find the population of the fundamental mode to first-order
\begin{equation}
    N_\mathrm{FSR} = \expval{\a_0^\dagger(t=\infty)\a_0(t=\infty)} \approx \expval{\a_0^{(0)\dagger}\a_0^{(0)}} + 2\mathrm{Re}\left\{\expval{\a_0^{(0)\dagger}\a_0^{(1)}}\right\}
\end{equation}
In the following sections we will solve the two perturbation equations and find an expression for $N_\mathrm{FSR}$.

\subsection{Solving the perturbation equations}

\subsubsection{0th order perturbation}
The 0th order equation is solved by integration
\begin{equation}
    \a_j^{(0)}(t) = -i\frac{2}{\sqrt{\kappa_b}} \int_{t_0}^t \d t' \Omega_j(t') \b_\mathrm{in}(t')\exp\left[-i\Delta_{a,j} (t-t') - \frac{2}{\kappa_b}\int_{t'}^t\d \tau |\Omega_j(\tau)|^2\right]
\end{equation}
For the purposes in this work we will only consider a single-photon input, we can therefore replace $\b_\mathrm{in}$ with $\psi_\mathrm{in}$, and let $\a_j^{(0)}(t) \rightarrow a_j^{(0)}(t)$ so that the analysis from here is with complex numbers~\cite{gorshkov_universal_2007}.

We want to maximize the population of the fundamental mode, so we will use the 0th order exression to determine the shape of $\Omega$ that achieves this.
First we define
\begin{equation}
    \zeta(t',t) = \frac{2}{\sqrt{\kappa_b}}\Omega_0(t')\exp\left[ - \frac{2}{\kappa_b}\int_{t'}^t\d \tau |\Omega_0(\tau)|^2\right]
\end{equation}
such that,
\begin{equation}
    a_0^{(0)}(t)= -i\int_{t_0}^t \d t' \zeta(t',t) \psi_\mathrm{in}(t')\,.
\end{equation}
To find the optimal pump-pulse shape we choose the shape that maximizes the field in the cavity at $t=T$ (where we interpret $T \rightarrow \infty$), i.e. $\zeta(t,T) = \psi_\mathrm{in}^*(t)$, such that:
\begin{equation}
    a_0^{(0)}(T)= -i\int_{t_0}^T \d t' |\psi_\mathrm{in}(t')|^2 = -i\,, \quad \text{i.e.:} \quad \int_{t_0}^T \d t' |\zeta(t',T)|^2 = 1
\end{equation}
Now we compute the integral over $\zeta$
\begin{equation}
\begin{aligned}
    &\int_{t_0}^t \d t' |\zeta(t',T)|^2 = \frac{4}{\kappa_b}\int_{t_0}^t\d t' |\Omega(t')|^2 \exp\left[\frac{4}{\kappa_b}\int_T^{t'}\d\tau |\Omega(\tau)|^2\right]\\
    =& \exp\left[-\frac{4}{\kappa_b}\int_t^T\d\tau  |\Omega(\tau)|^2\right] - \exp\Bigg[-\frac{4}{\kappa_b}\underbrace{\int_{t_0}^T\d\tau  |\Omega(\tau)|^2}_{=\infty}\Bigg] = \exp\left[-\frac{4}{\kappa_b}\int_t^T\d\tau |\Omega(\tau)|^2\right]
\end{aligned}
\end{equation}
where we have chosen the pulses given by $\Omega$ to diverge when integrated over the entire pulse length, such that $\zeta$ is normalized for $t=T$ as required.
Thereby we end up with
\begin{equation}
    \int_{t_0}^t \d t' |\zeta(t',T)|^2 = \int_{t_0}^t \d t' |\psi_\mathrm{in}(t')|^2= \exp\left[-\frac{4}{\kappa_b}\int_t^T\d\tau |\Omega(\tau)|^2\right]
\end{equation}
Using our requirement $\psi_\mathrm{in}^*(t) = \zeta(t,T)$ and the last equality sign in the above equation, we can solve for $\Omega$ and find
\begin{equation}
    \Omega(t) = \frac{\sqrt{\kappa_b}}{2}\frac{\psi_\mathrm{in}^*(t)}{\sqrt{\int_{t_0}^t\d t' |\psi_\mathrm{in}(t')|^2}}
\end{equation}
We note that this matches the result in the main text, given by Eq. (9), in the limit $\Tin \gg \Trt$, which is equivalent to the limit considered here.

We can now remove the $\Omega$-dependency from $a_j^{(0)}$ and thereby we find an expression for the cavity field-modes that only depends on the input and the free-spectral range $\FSR$
\begin{equation}
    a_j^{(0)}(t) = -i \left(\int_{t_0}^t\d t' |\psi_\mathrm{in}(t')|^2\right)^{-\frac{1}{2}}\int_{t_0}^t \d t' |\psi_\mathrm{in}(t')|^2\exp\left[-i\Delta_{a,j} (t-t') \right]
\end{equation}

\subsubsection{1st order perturbation}
Inserting the 0th order solution into the equation for the 1st order perturbation, together with the pump shape found before, we can integrate the equation and find the solution
\begin{equation}
    a_0^{(1)}(t) = \frac{i}{2} \int_{t_0}^t \d \tau \frac{\psi_\mathrm{in}^{*2}(\tau)}{\left(\int_{t_0}^\tau\d t' |\psi_\mathrm{in}(t')|^2\right)^{\frac{3}{2}}} \sum_{k\neq 0}  \int_{t_0}^{\tau} \d\tau'|\psi_\mathrm{in}(\tau')|^2\e^{-i\Delta_{a,k}(\tau-\tau')}
\end{equation}
To simplify this result, we assume $\Delta_{a,k}$ is much faster than the variations in $\psi_\mathrm{in}$, so we can pull the integral in the denominator out of the $\tau$ integral and thereby get
\begin{equation}
    a_0^{(1)}(t) \approx \frac{i}{2} \left(\int_{t_0}^t\d t' |\psi_\mathrm{in}(t')|^2\right)^{-\frac{3}{2}}\int_{t_0}^t \d \tau 
    \psi_\mathrm{in}^{*2}(\tau) \sum_{k\neq 0}  \int_{t_0}^{\tau} \d\tau' |\psi_\mathrm{in}(\tau')|^2 \e^{-i\Delta_{a,k}(\tau - \tau')}
\end{equation}

\subsection{Fundamental mode population}
Using the results from the previous section we can express the population of the fundamental mode as
\begin{equation}
    N_\mathrm{FSR} \approx 1 - \left(\int_{t_0}^t\d t' |\psi_\mathrm{in}(t')|^2\right)^{-\frac{3}{2}} \mathrm{Re}\left\{\int_{t_0}^t \d \tau\, 
    \psi_\mathrm{in}^{*2}(\tau) \sum_{k\neq 0}  \int_{t_0}^{\tau} \d\tau' |\psi_\mathrm{in}(\tau')|^2 \e^{-i\Delta_{a,k}(\tau - \tau')}\right\}
\end{equation}
We note here that this resembles, but is not quite the same as the frequency component at the relevant sideband frequency. 
The reason is that the drive strength  depends on the time, resulting rather in the Fourier component of the square of the field at the sideband frequency (for pulses without a time dependent phase).
In the following we present the results of these integrals for two different relevant input mode functions.

\subsubsection{Exponentially decaying input}
We start by considering an exponential decay, such as the spontaneous emission from a quantum dot, which is described by
\begin{equation}
     \psi_\mathrm{in}(t) = \sqrt{\kappa_\mathrm{in}} \Theta(t-t_0)\e^{-\frac{1}{2}\kappa_\mathrm{in}(t-t_0)}
\end{equation}
where $\Theta$ is the Heaviside function, and $\Tin = \kin^{-1}$ is the standard deviation of $|\psi_\mathrm{in}|^2$.
With this input we find:
\begin{equation}
    N_\mathrm{FSR}= 1 - \frac{1}{2}\sum_{k\neq 0}\frac{\kappa_\mathrm{in}^2}{\Delta_{a,k}^2 + \kappa_\mathrm{in}^2} = 1 - \frac{1}{2}\left[\pi\frac{\kappa_\mathrm{in}}{\FSR}\coth\left(\pi\frac{\kappa_\mathrm{in}}{\FSR}\right) - 1\right]
\end{equation}
The sum represents the number of sidebands we include.
In \cref{fig:appendix} we have illustrated how the number of sidebands impacts the efficiency and compared it with the exact summation.

\subsubsection{Gaussian input}
\begin{figure}[h]
    \centering
    \includegraphics[width=\textwidth]{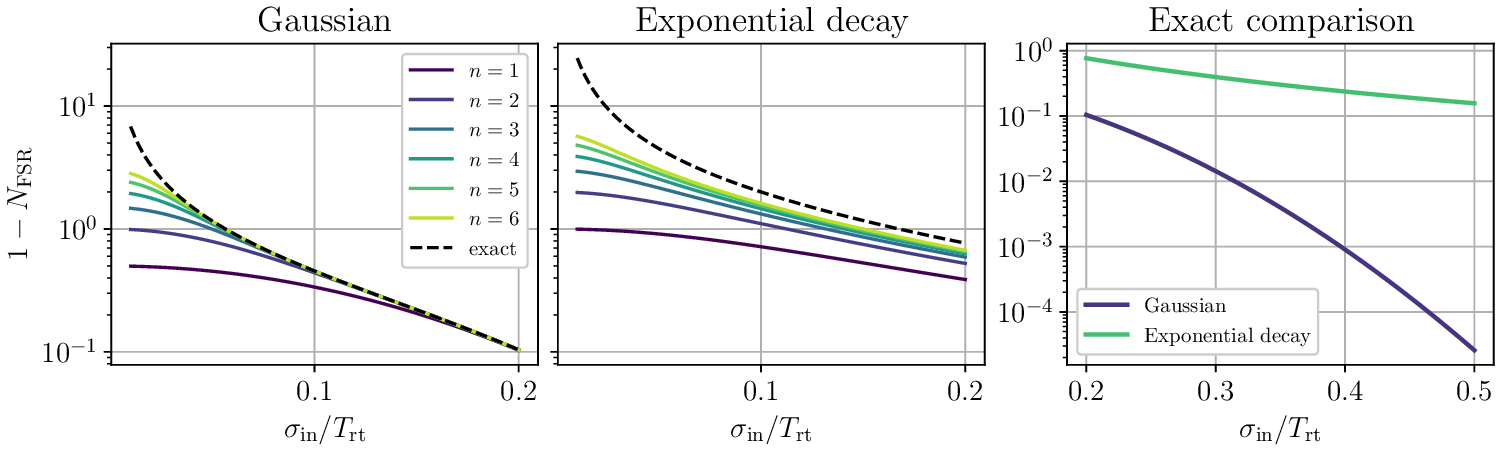}
    \caption{The approximated stretching efficiency for a Gaussian and an exponentially decaying input. The exact summations are compared to the partial summations using $n$-number of sidebands on each side of the fundamental mode. In the plot to the right, the exact summations of the Gaussian and Lorentzian inputs are compared for broader pulses. As evident in the plot, the Gaussian pulse has a much faster convergence to unit efficiency since the Gaussian is much more narrow in frequency. }
    \label{fig:appendix}
\end{figure}

We now consider a Gaussian input, as used in the maintext
\begin{equation}
    \psi_\mathrm{in}(t) = \left(\frac{\kin^2}{2\pi} \right)^{\frac{1}{4}} \e^{-\left(\frac{\kappa_\mathrm{in} t}{2}\right)^2}
\end{equation}
where $\Tin = \kin^{-1}$ is the standard deviation of $|\psi_\mathrm{in}|^2$.
With this input we find:
\begin{equation}
    N_\mathrm{FSR} = 1 - \frac{1}{4}\sum_{k\neq 0} \e^{-\left(\frac{\Delta_{a,k}}{\kappa_\mathrm{in}}\right)^2} = 1 - \frac{1}{4}\left[ \Theta_3\left(0,\e^{-\left(\frac{\FSR}{\kappa_\mathrm{in}}\right)^2}\right) -1 \right]
\end{equation}
where $\Theta_3(a,b)$ is the Jacobi-Theta function of 3rd order.
Again the sum in the middle expression represents the number of sidebands included, where the contribution of the first sideband on each side of the fundamental mode amounts to $N_\mathrm{FSR} = 1 - \frac{1}{2}\e^{-\left(\frac{\FSR}{\kappa_\mathrm{in}}\right)^2}$, as used in the main text.
In \cref{fig:appendix} the impact of the sidebands are shown and compared to the exact summation.
For $\Tin/\Trt > 0.2$ the first sideband approximation agrees well with the exact summation.

\end{document}